# Coherence and Phase in an Electronic Mach-Zehnder Interferometer: An Unexpected Behavior of Interfering Electrons

I. Neder, M. Heiblum, Y. Levinson, D. Mahalu, and V. Umansky

*Braun Center for Submicron Research, Dept. of Condensed Matter Physics,*

*Weizmann Institute of Science, Rehovot 76100, Israel*

**We report the observation of an unpredicted behavior of interfering 2D electrons in the integer quantum Hall effect (IQHE) regime[1] via a utilization of an electronic analog[2] of the well-known Mach-Zehnder interferometer (MZI)[3]. The beauty of this experiment lies in the simplicity of two path interference. Electrons that travel the two paths via edge channels, feel only the edge potential and the strong magnetic field; both typical in the IQHE regime. Yet, the interference of these electrons via the Aharonov-Bohm (AB) effect[4], behaves surprisingly in a most uncommon way. We found, at filling factors 1 and 2, high visibility interference oscillations, which were strongly modulated by a lobe-type structure as we increased the electron injection voltage. The visibility went through a few maxima and zeros in between, with the phase of the AB oscillations staying constant throughout each lobe and slipping abruptly by π at each zero. The lobe pattern and the 'stick-slip' behavior of the phase were insensitive to details of the interferometer structure; but highly sensitive to magnetic field. The observed periodicity defines a 'new energy scale' with an unclear origin. The phase rigidity, on the other hand, is surprising since Onsager relations are not relevant here[5,6].**

Under strong magnetic fields, in the IQHE regime, 2D electrons propagate in a beam like motion, performing chiral skipping orbits along the edges of the sample. Hence, forward and backward moving electrons propagate along opposite edges, quenching backscattering. Quantizing this motion leads to chiral edge states propagating in 1D like edge channels. The electrons thus move ballistically, with extremely long scattering lengths, enabling constructing relatively large interferometers[1]. An electronic two path MZI is described in Fig. 1. It is based on a 2D electron gas that is formed in a GaAs-AlGaAs heterostructure, some 60nm below the surface. Instead of the beam splitter in the optical MZI, which is used either to split the incident photon beam to two paths or to interfere the two paths later, we employed partly transmitting potential barriers formed by quantum point contacts (**QPC**s). An Ohmic contact served as source **S2** to inject electrons and two other contacts served as drains **D1** and **D2** to collect the outgoing currents. A modulation gate **MG** controlled the phase difference between the two paths $\varphi$, by changing the enclosed area between the two paths, leading to current oscillation in **D1** and in **D2** (in opposite phase). The period of the oscillations in units of area is $\Delta A = h/eB$, with $B$ the magnetic field, $h$ the Planck constant, and $e$ the electron charge.

In the linear regime (without DC bias applied) the conductance from source **S2** to drain **D1** or **D2** corresponds to transmission probability $T_{SD}$ via the Landauer and Buttiker formalizm[7]. The transmission, in turn, depends on the interference between the amplitudes of the two paths. When the system is tuned to the IQHE regime, say filling factor $\nu=1$, a single edge current is injected from **S2**, splits to a transmitted and reflected edge channels by **QPC1**, with both propagating along the inner and outer boundaries of the ring. For transmission and reflection amplitudes $t_i$, $r_i$ of the $i^{th}$ QPC with $|r_i|^2+|t_i|^2=1$; the collected currents at **D1** and **D2** can be expressed as: $I_{D1} \propto T_{S2D1}= |t_1 r_2+r_1 t_2 e^{i\varphi}|^2 = |t_1 r_2|^2 + |r_1 t_2|^2 - 2|t_1 t_2 r_1 r_2|\cos\varphi$, and $I_{D2} \propto T_{S2D2}= |t_1 t_2+r_1 r_2 e^{i\varphi}|^2 = |t_1 t_2|^2 + |r_1 r_2|^2 + 2|t_1 t_2 r_1 r_2|\cos\varphi$ and obviously $I_1+I_2=I_{S2}$. We define the visibility of the oscillating currents in the drains $v=(I_{max}-I_{min})/(I_{max}+I_{min})$.

The new MZI was similar to its predecessor[1] except for an additional source **S3** and a quantum dot (QD) - from it only **QPC0** was operated - inserted in the current path to **QPC1** (see Fig. 1). The additional source enabled injection of electrons with an arbitrary chemical potential, whiles **QPC0** allowed selective reflection of edge channels in ν=2 or dilution of the impinging current. Differential measurements were performed by superimposing a small AC voltage (~1μV at ~1MHz) on a DC bias at **S2**, and measuring the AC voltage at **D2** at an electron temperature ~20mK. The AC signal was amplified by an in situ, home made, preamplifier cooled to 4.2K[1], followed by a room temperature amplifier and a spectrum analyzer. The relative phase between the two paths was modulated by a voltage $V_{MG}$ (on **MG**). As will be shown below, neglecting decoherence a single particle model predicts the differential visibility of the AB oscillations to be energy independent.

Previous measurements, performed with an electronic MZI but without employing **QPC0** after the source[1], resulted with a smooth decay of the visibility with applied DC voltage. The visibility also dropped in a similar fashion when the temperature was increased to ~100mK at $V_{DC}=0$. Shot noise measurements suggested that the rapid loss of interference did not result from decoherence. A recent paper suggested that some low frequency fluctuations, such as $1/f$ noise, change randomly the area of the MZI and lead to quenching of the visibility[8]. We show here a much richer set of experimental results that puts in doubt the above models.

Most of the measurements were conducted at $\nu=2$, with two edge channels injected from **S2**, since the device seemed to be more stable at lower magnetic fields. However, unlike reported in Ref. 1, the inner edge was now totally reflected by **QPC0** and only the outer edge channel arrived at the MZI. For zero DC bias on **S2** and $T_{QPC1}=T_{QPC2}=1/2$, the maximum visibility measured at **D2** was ~60%. In Fig. 2a we present a two-dimensional color plot of the AC voltage measured at **D2** as function of $V_{MG}$ and the applied DC bias for $\nu=2$. Figures 2b ($\nu=1$) and 2c ($\nu=2$) provide the normalized amplitude (visibility) of and phase of the AB oscillations at different values of $V_{DS}$ - derived from such plots as in Fig. 2a (via performing a fast complex Fourier transform on the $V_{MG}$ axis). Two striking features are common to Figs. 2b and 2c: (a) The visibility evolves in a form of a decaying lobe pattern with increasing $V_{DC}$; dipping to zero periodically at specific biasing voltages, and (b) The phase is constant throughout each lobe but slips abruptly by $+\pi$ or $-\pi$ at each zero. It is apparent that at $\nu=2$ there are more pronounced lobes than at $\nu=1$. This surprising behavior of a beating visibility and a rigid phase presents a new energy scale of ~10µeV. Moreover, as the magnetic field was reduced the amplitude and the periodicity of the lobe pattern got smaller and eventually vanished at magnetic fields in the lower half of the $\nu=2$ plateau (see Fig. 3).

Figures 2 and 3 summarized the main results of this work. The results clearly deviate from the single particle model in the linear regime. We look now in more detail on the system under study. The probed chemical potential at **D2** is a mixture of the chemical potentials of the two edge channels emanating from **S1** (acts also as **D1**) and **S2**. Hence, the chemical potential of **D2** can be written as,

$$\mu_{D2} = \overline{T}\mu_{S2} + (1-\overline{T})\mu_{S1} , \tag{1}$$

where $\mu_{S1}$ and $\mu_{S2}$ are the chemical potentials of the two sources, $\overline{T}$ is the average transmission from **S2** to **D2**. The transmission $\overline{T}$ depends on the individual transmissions through **QPC1** and **QPC2**, on the AB phase, and in the most general case also on $\Delta\mu \equiv \mu_{S2} - \mu_{S1}$. Equation 1 can be also written in a more familiar manner: $\mu_{D2} = \mu_{S1} + \overline{T}\Delta\mu$. Superimposing a small AC signal on a DC bias at **S2** and measuring only the AC response at **D2** (being a measure of the derivative), leads to:

$$\delta V_{D2}^{AC} = \left(\frac{d\mu_{D2}}{d\mu_{S2}}\right)_{eV_{DC}} \delta V_{S2}^{AC} . \tag{2}$$

Using Landauer-Buttiker formalism at zero temperature, one integrates over the single particle energy dependent transmission coefficient in order to get the chemical potential at the drain,

$$\mu_{D2} = \mu_{S1} + \int_{\mu_{S1}}^{\mu_{S2}} T_{S2D2}(\varepsilon, B, V_{MG}) d\varepsilon . \tag{3}$$

Since the transmission of each QPC is fairly constant with energy ε, the dependence of $T_{S2D2}$ on energy is mostly through its dependence on the phase of each single particle state. In turn, the dependence of the phase is given by:

$$\varphi = \varphi_0(B) + \beta V_{MG} + \frac{\Delta L}{\hbar v_g}(\varepsilon - \varepsilon_f) \; , \qquad (4)$$

with $\varphi_0(B)$ the AB phase at the Fermi energy $\varepsilon_f$, $\beta V_{MG}$ is the added phase due to the modulation gate voltage, $\Delta L$ is the length difference between the two path, and $v_g$ is the single particle group velocity at energy $\varepsilon_f$ (assumed to be constant for small enough DC excitation voltage). From Eqs. 2 and 3 we find an energy independent 'differential visibility' $\frac{2\sqrt{t_1 r_1 t_2 r_2}}{t_1^2 t_2^2 + r_1^2 r_2^2}$, with the phase of the AB oscillations proportional to the applied DC voltage at **S2**, $\Delta\varphi \propto \frac{\Delta L}{\hbar v_g} V_{DC}$. This clearly contradicts the experimental results.

However, one can notice in the above discussion that the single particle picture is inconsistent when the transmission is energy dependent. Look at the term $\bar{T} = \frac{1}{\Delta\mu} \int_{\mu_{S1}}^{\mu_{S2}} T_{SD}(\varepsilon) d\varepsilon$. It must be invariant to an addition of potential $V$ to both sources but be sensitive only to the difference $\Delta\mu$. However, the term $\int_{\mu_{S1+V}}^{\mu_{S2+V}} T_{SD}(\varepsilon) d\varepsilon = \int_{\mu_{S1}}^{\mu_{S2}} T_{SD}(\varepsilon + V) d\varepsilon$ depends on $V$ and therefore cannot describe properly the real system. Hence, in order to make the expression consistent 'self charging' of each edge must take place, with each single particle state that corresponds to wave number $k$ having now energy $\varepsilon(k)+eV$ instead of $\varepsilon(k)$. This is a mean field approximation since it ignores charge fluctuations at each edge. Note that we implicitly assumed here that the capacitance of each edge channel is $c=Le^2/hv_g$. Employing now this self consistent approach, the highest occupied state has energy $\varepsilon_f + T_{QPC1}\Delta\mu$ in one arm and $\varepsilon_f + R_{QPC1}\Delta\mu$ in the other. This leads to a shift of the energy of each state by a new average chemical potential and the expression for the phase in Eq. 4 is now modified:

$$\varphi = \varphi_0(B) + \beta V_{MG} + \frac{L_1}{\hbar v_g}\left[\varepsilon - \left(\varepsilon_f + T_{QPC1}\Delta\mu\right)\right] - \frac{L_2}{\hbar v_g}\left[\varepsilon - \left(\varepsilon_f + R_{QPC1}\Delta\mu\right)\right] \quad . \quad (5)$$

Plugging this phase back into $T_{S2D2}$ in Eq. 3 solves the inconsistency, leading to an integrant that is bias dependent. The differential response (defined in Eq. 2) now depends on the whole energy range (determined by the applied voltage). For the special condition $T_{QPC1}=R_{QPC1}=0.5$, we find:

$$\left(\frac{d\mu_{D2}}{d\mu_{S2}}\right)_{eV_{DC}} = 0.5 + \sqrt{T_2 R_2} \cdot \cos\left[\Phi_0(B) + \beta V_{MG}\right] \cdot \cos\left(\frac{\Delta L}{\hbar v_g}eV_{DC}\right) \quad . \quad (6)$$

Equation 6 describes AB oscillations with constant phase and with amplitude multiplied by a cosine term with an argument dependent on the applied bias. The zeros of the visibility, where the π phase lapses in the experiment took place, are at $\frac{\Delta L}{\hbar v_g}eV_{DC} = \frac{\pi}{2} + n\pi$, defining a periodicity that might be related to some energy scale.

Equation 6 looks promising since it provides an explanation for the lobe structure and the rigidity of the phase, however, it is highly sensitive to the condition $T_{QPC1}=R_{QPC1}=0.5$ and to the paths length difference $\Delta L$. Deforming the symmetry of the interferometer, either via varying the transmission of **QPC1** or by changing $\Delta L$ provides a test of the above hypothesis. In the first measurement we varied $T_{QPC1}$ from 0.1 to 0.9, hence charging the two arms differently. However, except for a change in the overall magnitude of the visibility, which followed a $\sqrt{T_{QPC1}(1-T_{QPC1})}$ dependence, the lobe structure and the phase rigidity remained as before. In the second measurement we increased the length of one path in two equal steps (via removing the negative voltage from gates **G1** and **G2**, seen in Fig. 1). As seen in Fig. 4, whiles the amplitude of the visibility decreased with path lengthening the overall lobe pattern retained the same periodicity and the phase remained unaffected.

While these two tests affected the geometry of the interferometer, their effect on the self consistent electro chemical potential of each path can be rather complicated. However, by utilizing the inner edge channel (at ν=2) we can capacitively charge the outer edge. This can be done by injecting also from source **S3** and tuning **QPC0** so that to the MZI two edges arrive: the outer from **S2** and the inner from **S3** (Fig. 1). Consequently, inside the MZI the two edges followed the same path, passing under the metal bridge of **D1** - thus mutually interacting along path length of some 10μm. We applied DC bias to **S3** while keeping the outer edge, injected by **S2**, at $V_{DC}$=0 and a small AC signal. As seen in Fig. 5a, the phase of the AB oscillations followed this time a linear dependence on the DC bias of **S3**, clearly showing that the inner edge channel functioned as a modulating gate, adding phase to the inner edge, with an extremely high phase sensitivity of ~2π/30μV. This is a clear demonstration of strong inter edge electron-electron interaction. Hence, it is likely that the intra edge interactions are also strong leading to self charging.

How does such intentional charging of the inner edge affect the lobe pattern? This was studied by injecting the two edge channels by **S2**, biased DC+AC, and fully opening **QPC0**. This experiment is similar to that reported before in Ref. 1. As then, and seen in Fig. 5b, the lobe structure disappeared and the visibility dropped nearly monotonically as function as $V_{DC}$. Moreover, the phase did not exhibit any more a 'stick-slip' type behavior, but a smooth change that is linear with $V_{DC}$ at low voltage. These measurement results are not independent: they could, in principle, be predicted from the data in Fig. 5a and Fig. 2; hence they do not add much to our understanding aside from confirming the older data[1].

We chose to present here one more (confusing) result by studying again intra edge interactions. A way to affect directly the charging of both paths simultaneously without altering the energy of the injected particles is via dilution of the incoming beam. This can be easily done by partly reflecting the outer, interfering, edge channel by **QPC0**. Note that the short distance the dilute edge channel must propagate to arrive at **QPC1** of the MZI is not sufficiently long to allow energy relaxation; hence the highest energy of the electrons arriving at **QPC1** is still $\mu_{S2}+eV_{DC}$. Figure 6 shows the dramatic effect of such dilution. The higher order lobes stretched out to higher voltages and weakened significantly, but the main lobe remained almost invariant.

By now it is obvious that the energy dependence of the two path interference in this, almost ideal, interferometer was not expected. An important question is whether this far-from-the-linear regime behavior reflects a general behavior of a system with a transmission that is strongly dependent on energy, or, is it specific to a system in the IQHE regime. In the first, more general, case it is a clear demonstration of the breakdown of Landauer's conductance picture. A strong dependence of the visibility on the applied DC bias was also observed before in a two-terminal two path interferometer without the influence of strong magnetic field[6]. There, the abrupt phase slips and the periodic behavior of the visibility were explained by invoking Onsager relations[7]; even though the system was not in the linear regime. Here, moreover, time reversal symmetry does not exist due to the high magnetic field applied, making this explanation of phase rigidity even more questionable.

We summarize now the main experimental results: (a) Evolution of the visibility with energy presented a surprising descending lobe pattern, which might have resulted from some type of phase averaging. The periodicity of the lobe structure, being 10-20μV in $V_{DC}$, presents a new and challenging 'energy scale'; (b) Strong inter edge interactions suggest also intra edge interactions; (c) Dilution of the propagating beams that is expected to affect intra edge electron interactions, indeed affected strongly the lobe pattern and the phase; (d) Tuning the interferometer away from symmetry had negligible effect on the visibility. The last 'fact' makes our proposed naïve model, which invokes short range electron interaction in the edge channel, invalid. One has to find a self consistent theory, which might depend on long range interactions (across the whole size of the interferometer) that will account for this new interference effects[9].

## Acknowledgement

We are indebt to Y. Chung, and F. Portier and A. Ra'anan for their valuable help in the design, fabrication and measurements, and to M. Buttiker for helpful discussions. The work was partly supported by the MINERVA foundation, the German Israeli Project Cooperation (DIP) and the GIF foundation.

# Figure Captions

**Figure1.** A top SEM micrograph of the Mach Zehnder Interferometer (MZI). The edges of the sample were defined by plasma etching of the GaAs-AlGaAs heterostructure, which embeds a high mobility 2D electron gas, some 80nm below the surface. Edge channels were formed by applying a perpendicular magnetic field, with filling factors $\nu=1$ or 2 in the bulk. The quantum dot that follows the source **S2** (not seen in the picture) was not used, but one of its quantum point contacts (**QPC0**) served to reflect back or to partly dilute the desired edge channel. The incoming edge channel (red line) was split by **QPC1** to two paths (dotted red lines), of which one moved along the outer edge of the device and the other around the inner drain **D1** (under the metallic air bridge). The two paths met again at **QPC2**, interfered, and resulted in two complementary currents: one in **D1** and the other in **D2**. The modulation gate (**MG**) changed the contour of one path, thus changing the enclosed area between the two paths and the phase difference between them (via the Aharonov-Bohm effect). Significant changes in path length could also be done by opening gates **G1** and **G2**. The signal at **D2** was filtered around a center frequency of ~1MHz with a cold LC resonant circuit and than was amplified by a low noise, preamplifier, cooled to 4.2K. Note that the centrally located small Ohmic contact ($3\times3\mu m^2$) served both as **D1** and **S1**. Being grounded, via a long metallic air-bridge, it collected the interfering current while injecting edge states at zero chemical potential. Two smaller metallic air-bridges applied voltage to the inner gates of **QPC1** and **QPC2**. Another source, **S3** (not seen in the picture) was used to inject current along the upper-outer edge of the mesa, with an arbitrary chemical potential.

**Figure 2.** Interference oscillations and the visibility. **a**. Two dimensional color plot of the ~1MHz AC signal measured at **D2** as function of the applied DC bias at the source **S2** (that was biased by both 1MHz AC signal of ~1μV and DC) and the modulation gate voltage at filling factor ν=2 and **QPC1** and **QPC2** with transmission T~0.5. **b.** The visibility (defined as [$V_{D2}$(max)-$V_{D2}$(min)]/[$V_{D2}$(max)+$V_{D2}$(min)], with $V_{D2}$ the measured AC signal at **D2**) and the phase of the interference pattern at ν=2 as a function of the applied DC bias to **S2** (deduced from Fig. 2**a**). Five major lobes are visible, each ~14μV wide. The phase at each lobe is constant but it slips abruptly by π at each node. **c.** A similar graph to that of Fig. 2**b**, but at ν=1, exhibiting only 3 major lobes with similar 'stick-slip' phase behavior.

**Figure 3.** Magnetic field dependence of the visibility plotted as function of the applied $V_{DC}$. The magnetic field is chosen to be on the conductance plateau of filling factor ν=2. As the magnetic field weakened the periodicity of the lobe structure got smaller and the strength of the oscillations diminished.

**Figure 4.** The effect of changing the length of one path on the visibility. The length of the path ($L_1$) was increased in two steps by unbiasing two gates (**G1, G2**) that confine the outer edge of the interferometer. As $L_1$ changed from 8.8μm to 11.2μm the lobe periodicity and the phase behavior (not shown) did not change but the visibility got smaller.

**Figure 5.** The effect of a biased inner edge channel at $\nu=2$ on the interference. **a.** The outer, interfering, channel was biased only with small AC signal ($V_{DC}=0$) while an inner channel was injected from source **S3**, which was biased separately with a DC bias only. The inner channel followed the interfering channel along its path and in close proximity, and served as a charging gate (see Fig. 1). While the visibility of the interference (not shown) stayed roughly constant the phase of the oscillations exhibited a linear dependence on the DC voltage of the inner channel. **b**. The visibility and phase as function of $V_{DC}$ applied to **S2**, with both channels emanated from **S2**. The inner channel was fully reflected by **QPC1**, hence, was noiseless. The lobe pattern almost fully vanished as the visibility dropped uniformly with $V_{DC}$ to zero. Moreover, the phase is no longer rigid and seems to have a mixed evolution, between linear and rigid dependences.

**Figure 6.** The effect of beam dilution on the lobe pattern of the visibility. Dilution of the outer channel was performed by reflecting part of the impinging channel by **QPC0**. When the occupation of the incoming edge channel was reduced to some 20% the higher order lobes stretched and vanished while the main lobe remained almost unaffected.

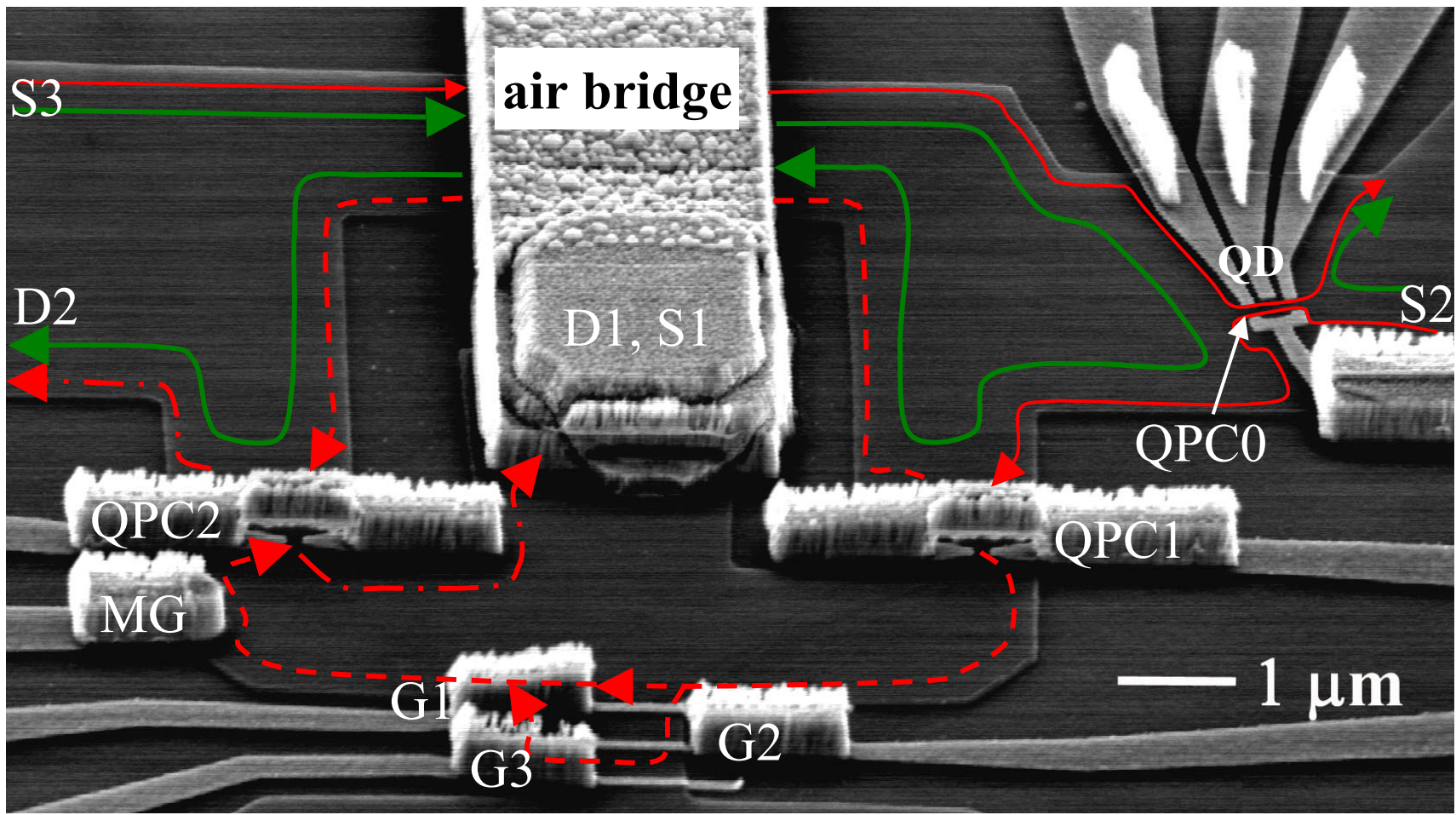

Fig. 1

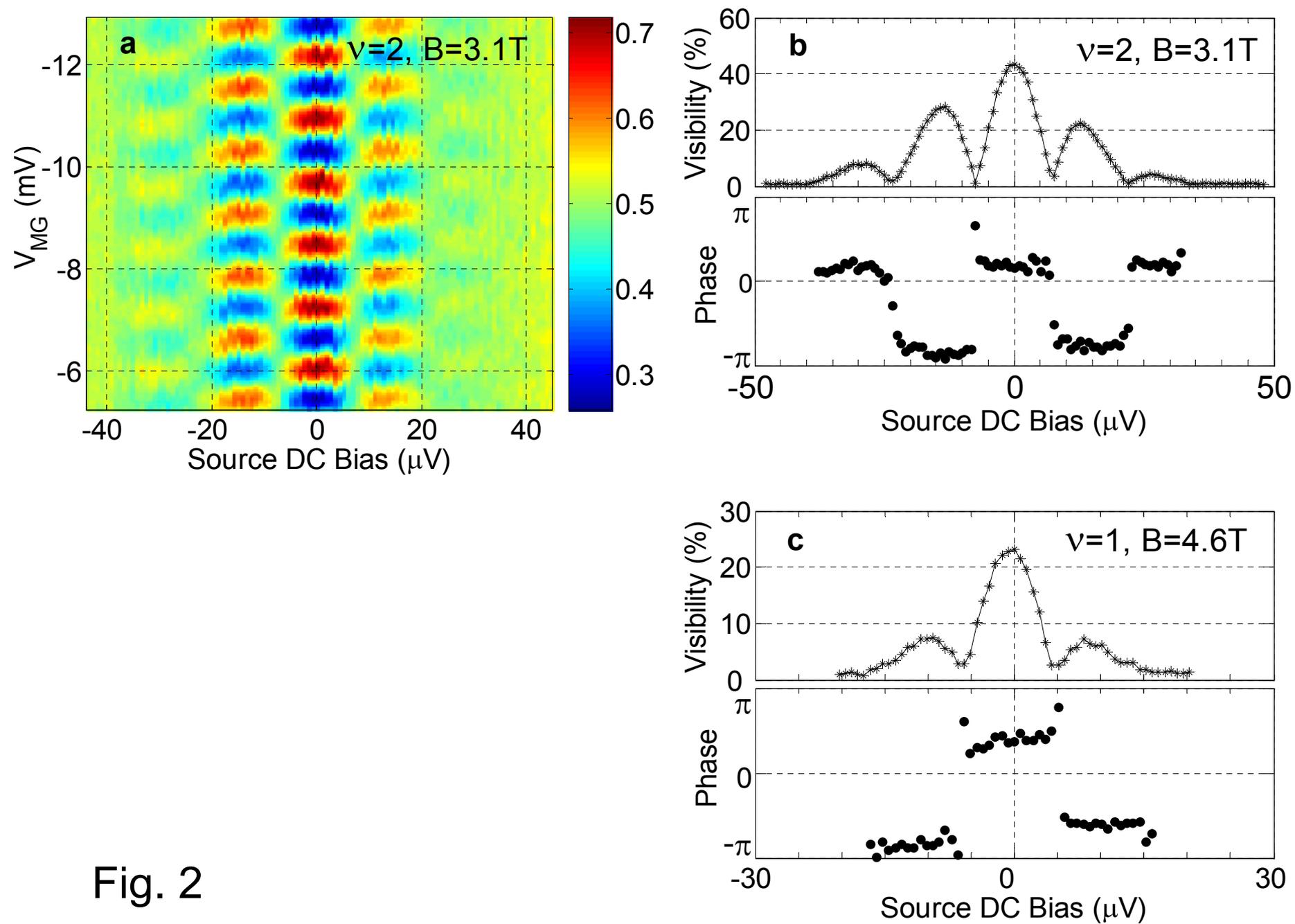

Fig. 2

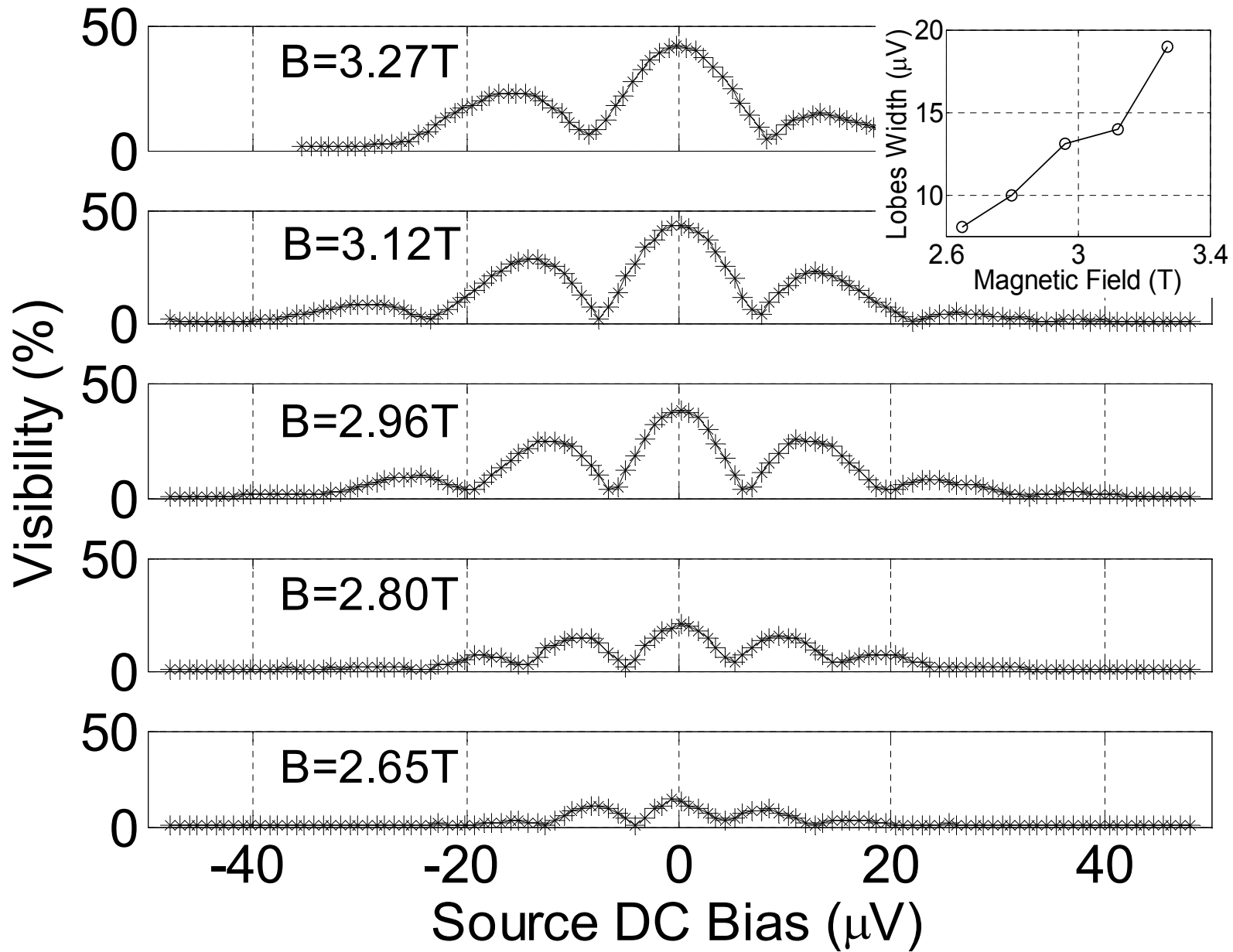

Fig. 3

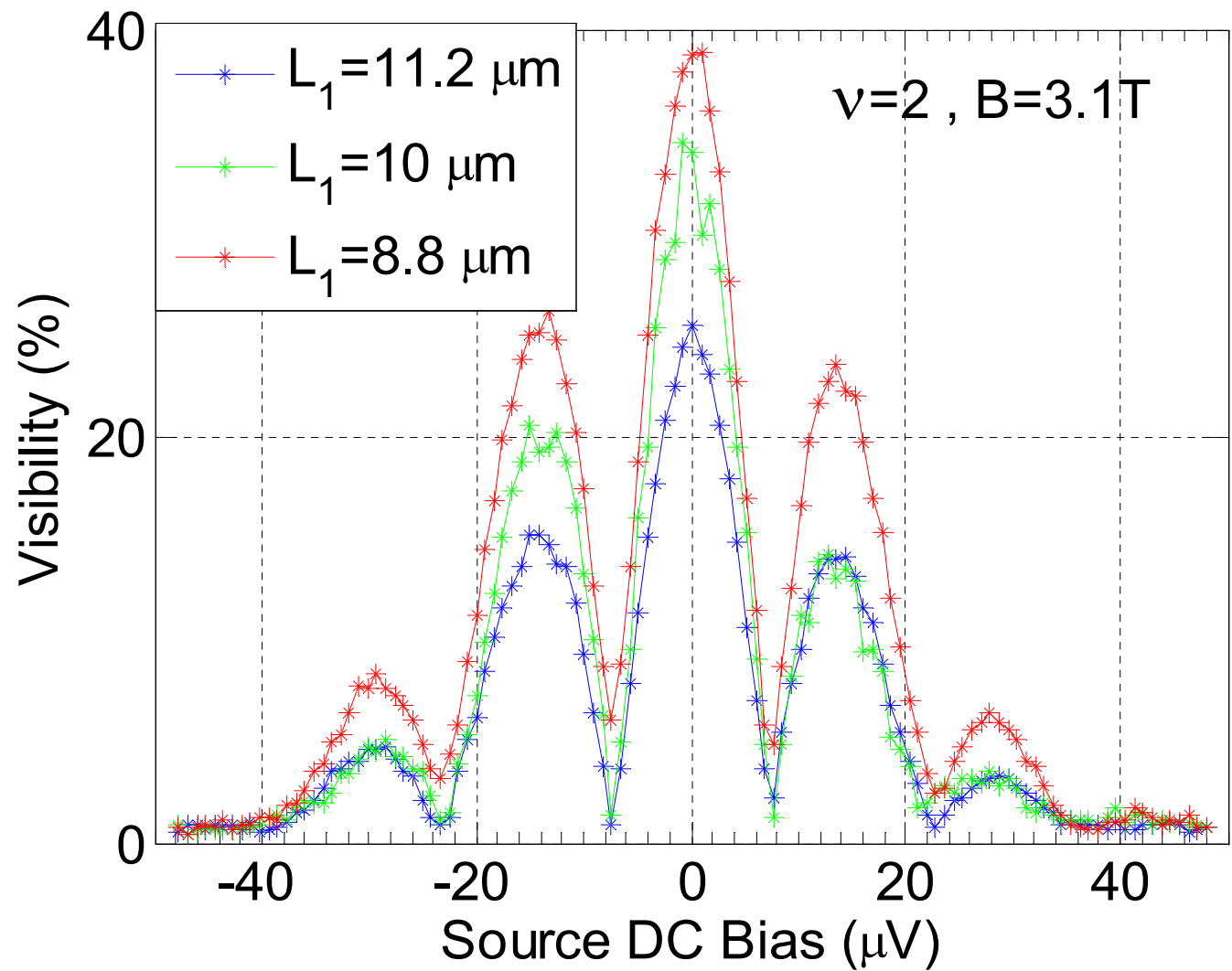

Fig. 4

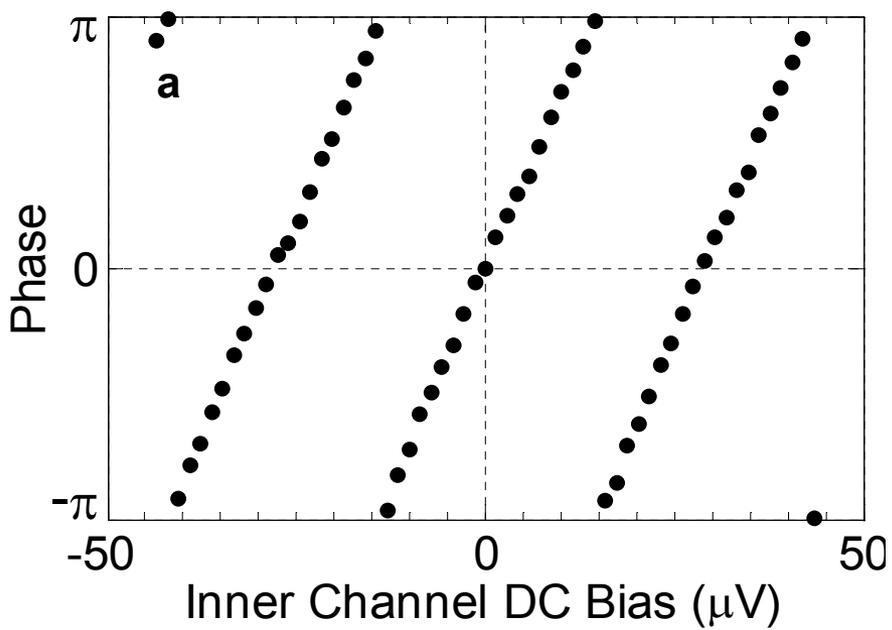 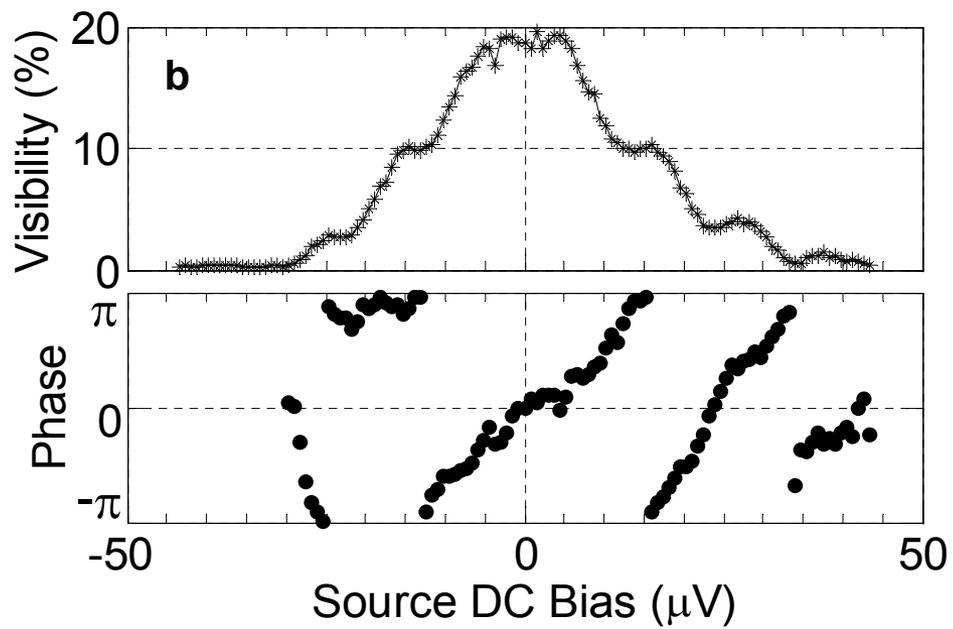

Fig. 5

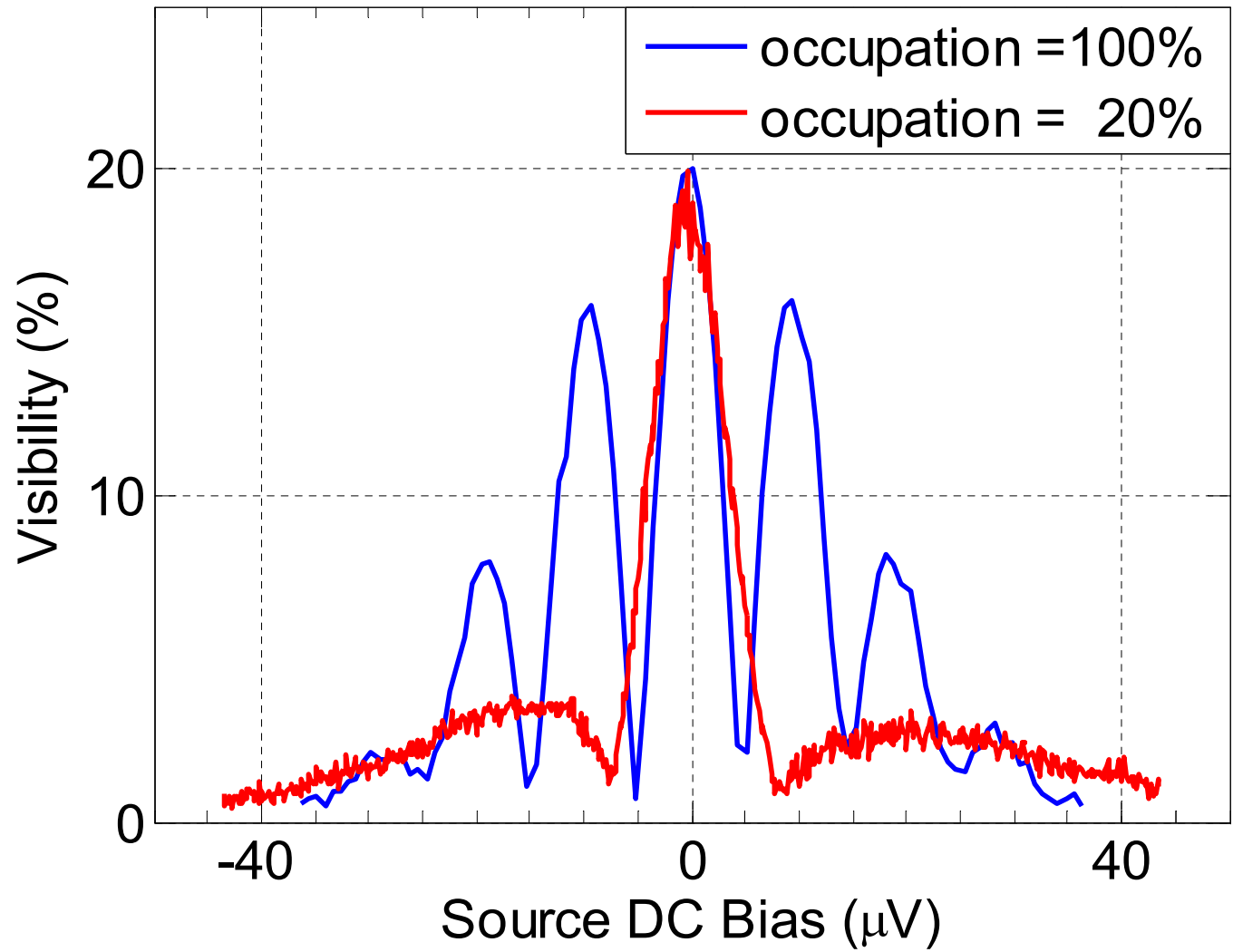

Fig. 6